\documentclass[groupedaddress,floatfix,showpacs,twocolumn]{revtex4}
\usepackage{graphicx}
\usepackage{wrapfig}
\usepackage{amsfonts}
\usepackage{epsfig}
\usepackage{epstopdf}
\usepackage{tabularx}
\usepackage{color}
\usepackage{setspace}
\begin{document}

\title{Nucleation and growth of a quasicrystalline monolayer: \\Bi adsorption on the five-fold surface of \emph{i}-Al$_{70}$Pd$_{21}$Mn$_{9}$}
\author{J.A.~Smerdon, J.K.~Parle, L.H.~Wearing}
\affiliation{Surface Science Research Centre and Department of Physics, The University of Liverpool, Liverpool L69 3BX, UK\\}
\author{T.A. Lograsso, A.R. Ross}
\affiliation{Ames Laboratory, Iowa State University, Ames, IA 50011, USA\\}
\author{R.~McGrath\footnote{Corresponding author. e-mail: mcgrath@liv.ac.uk; Ph.: +441517943364; Fax: +441517943444}}
\affiliation{Department of Physics and Surface Science Research Centre, The University of Liverpool, Liverpool L69 3BX, UK\\}
\pagenumbering{arabic}
\date{\today}

\begin{abstract}

Scanning tunnelling microscopy has been used to study the formation of a Bi monolayer deposited on the five-fold surface of \emph{i}-Al$_{70}$Pd$_{21}$Mn$_{9}$.  Upon deposition of low sub-monolayer coverages, the nucleation of pentagonal clusters of Bi adatoms of edge length 4.9 \AA{} is observed. The clusters have a common orientation leading to a film with five-fold symmetry.  By inspection of images where both the underlying surface and the Bi atoms are resolved, the pentagonal clusters are found to nucleate on pseudo-Mackay clusters truncated such that a Mn atom lies centrally in the surface plane. The density of these sites is sufficient to form a quasiperiodic framework, and subsequent adsorption of Bi atoms  ultimately leads to the formation of a quasicrystalline monolayer.  The initial nucleation site is different to that proposed on the basis of recent density functional theory calculations.

\end{abstract}

\maketitle

\section{Introduction} \label{intro}

Quasicrystals are metallic alloys with long-range order but without translational symmetry. Their surfaces have been studied extensively over the past decade. It has been found possible to prepare surfaces with a step/terrace morphology, and several experimental and theoretical investigations have led to a consensus that apart from a small degree of relaxation, the surfaces represent bulk terminations \cite{Fournee04,Sharma07}.

These surfaces provide interesting templates for the investigation of epitaxial growth. Such studies have been undertaken on several quasicrystals \cite{McGrath02-JPCM,Fournee04,Sharma07,Smerdon08-PM}, and quasicrystalline single element clusters and monolayers have been found on both icosahedral and decagonal quasicrystal surfaces. On the five-fold surface of icosahedral Al-Pd-Mn ($i$-Al-Pd-Mn), which is the substrate of interest in this work, Franke and co-workers found using low energy electron diffraction (LEED) and helium atom scattering (HAS) that Sb and Bi form ordered quasicrystalline monolayers \cite{Franke02}. Using STM, Si was found to order quasiperiodically at sub-monolayer coverages on $i$-Al-Pd-Mn  at the centres of truncated pseudo-Mackay clusters \cite{Ledieu06}. Recently using STM Pb was observed to grow in a quasiperiodic monolayer on $i$-Al-Pd-Mn where the building blocks are pentagonal clusters of edge length 4.9 \AA{} \cite{Ledieu08}. In that study the growth was found to proceed via self-assembly of an interconnected network of these pentagons, although the nucleation sites were not discussed.

In addition to the experimental work on the $i$-Al-Pd-Mn/Bi system described above \cite{Franke02}, there have been \emph{ab initio} DFT calculations of adsorption site energetics \cite{Krajci05}. Kraj\u{c}\'{\i}  and Hafner studied the adsorption of Bi on the 2/1 approximant to the $i$-Al-Pd-Mn quasicrystal \cite{Krajci05,Krajci06a}. This approximant is a periodic crystal whose unit cell contains 544 atoms and whose local structure is similar to that of the quasicrystal itself. A Penrose P1 tiling of edge-length 7.76 \AA{} was overlaid on the surface of the model as a guide to interpreting the model structure. This tiling consists of pentagons, rhombs, stars and boats \cite{Penrose74} (see Fig. \ref{3-2composite}(a)). The binding energies for Bi atoms on specific adsorption sites on the surface were calculated and then used for the initial positions of Bi atoms to investigate the stability of a full Bi monolayer.  The most favourable adsorption sites, in order of decreasing binding energy, were found to be at the centres of surface vacancies (which are contained within a subset of one orientation of the Penrose tiles); atop Mn atoms, (again in the centres of a subset of the pentagonal tiles); the vertices of the P1 tiling and the mid-edge positions of the tiling. The number of surface vacancies and of Mn atoms was considered to be too small for the formation of a quasiperiodic framework, and therefore it was proposed that the vertices and mid-edge positions of the tiling form the skeleton of a possible stable quasiperiodic monolayer. Additional atoms form smaller pentagons inside both orientations of pentagonal tiles. By allowing relaxation of the atomic forces the stability of the proposed monolayer structure was tested and predicted to have `pseudodecagonal symmetry'.

Given the prior experimental and theoretical work on this system, we have therefore investigated the initial growth of Bi on this five-fold surface using scanning tunnelling microscopy. The goals of this work were to test the predictions of the DFT calculations for the adsorption sites, and to gain insight on the growth mechanism for quasicrystalline monolayers.

\section{Experimental details} \label{exper}

The experiments were carried out in an Omicron variable temperature STM ultra high vacuum chamber.  The base pressure of the system was 1$\times$10$^{-10}$ mbar.   The \emph{i}-Al$_{70}$Pd$_{21}$Mn$_{9}$ quasicrystal sample, produced by the Ames group using the Bridgman method, was polished successively with 6 $\mu$m, 1 $\mu$m and 1/4 $\mu$m diamond paste before introduction to vacuum and thereafter was prepared in cycles consisting of 45 minutes sputtering with 3 keV Ar$^{+}$ ions followed by at least 4 hours annealing to 950 K, using electron-beam heating, up to a total annealing time of 20 hours. Bismuth was evaporated from a Mo crucible using an Omicron EFM-3 electron-beam evaporator with a flux monitor reading of 120 nA giving a deposition rate of approximately 1.8 ML per hour.  The chamber pressure did not exceed 2.5$\times$10$^{-10}$ mbar during evaporation.

\section{Results} \label{results}

\begin{figure}
\begin{center}
\includegraphics[width=0.4\textwidth]{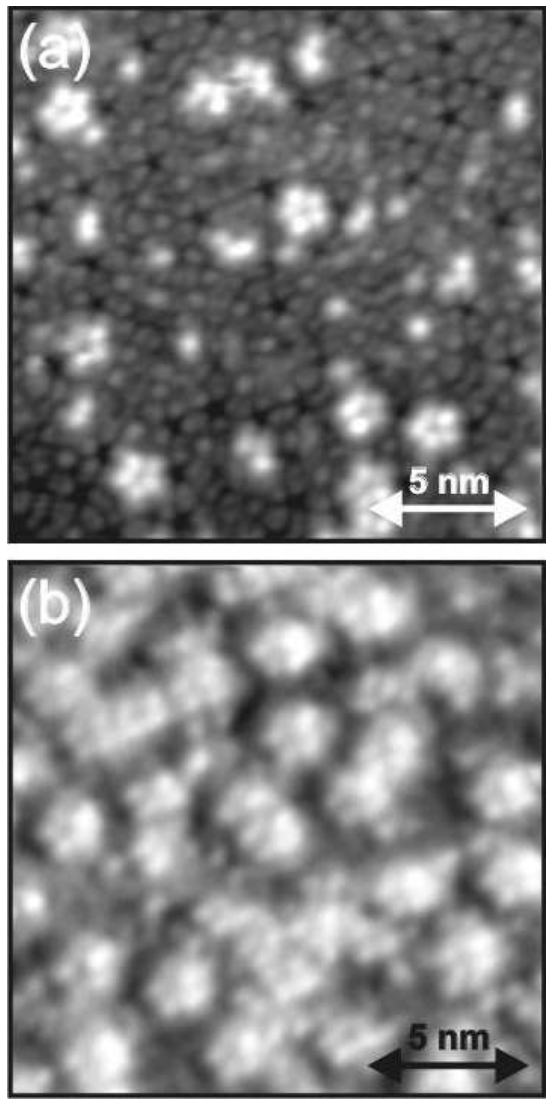}
\caption{Overview of the nucleation and growth of the Bi monolayer by room temperature deposition  on \emph{i}-Al-Pd-Mn. (\emph{a}); 0.13 ML Bi; (\emph{b}); 0.38 ML Bi.}
\label{bigrowth}
\end{center}
\end{figure}

The initial development of the Bi overlayer is followed in the STM images in Fig. \ref{bigrowth}. Fig. \ref{bigrowth}(a), taken at a Bi coverage of 0.13 ML, shows that growth proceeds through the nucleation of small clusters. In this nucleation process the stable clusters are small Bi pentagons, reflecting the symmetry of the substrate.  These pentagonal clusters can be seen to have a common orientation (leading to a five-fold symmetric submonolayer film). The distance between the centres of the bright protrusions forming these pentagons is 4.9 $\pm$ 0.2 \AA. The closest distance between the centres of Bi pentagons is 12.0 $\pm$ 0.2 \AA{}. The underlying surface structure is still very well resolved at this coverage, and this is the basis for the further analysis described in Section \ref{analysis}. A fast Fourier transform (FFT) of an image of dimensions 1500 \AA{} $\times$ 1500 \AA{} was taken at this coverage (not shown). Due to the large scale of the image, the clean surface structure was no longer resolved; nevertheless an  excellent degree of quasiperiodic order was observed in the FFT indicating that the pentagonal Bi clusters are positioned on sites which themselves are quasiperiodically arranged.

Fig. \ref{bigrowth}(b) shows the system after deposition of 0.38 ML. Due to the higher Bi coverage the resolution of the STM images is degraded and details of the clean surface are no longer visible. The pentagonal Bi clusters are still observable. Growth at this stage is still in the \emph{pure nucleation regime} \cite{Brune98}, where the island size does not change despite an increase in the coverage of a factor of three, indicating that additional deposition results in the formation of new nuclei.

\begin{figure}
\begin{center}
\includegraphics[width=0.4\textwidth]{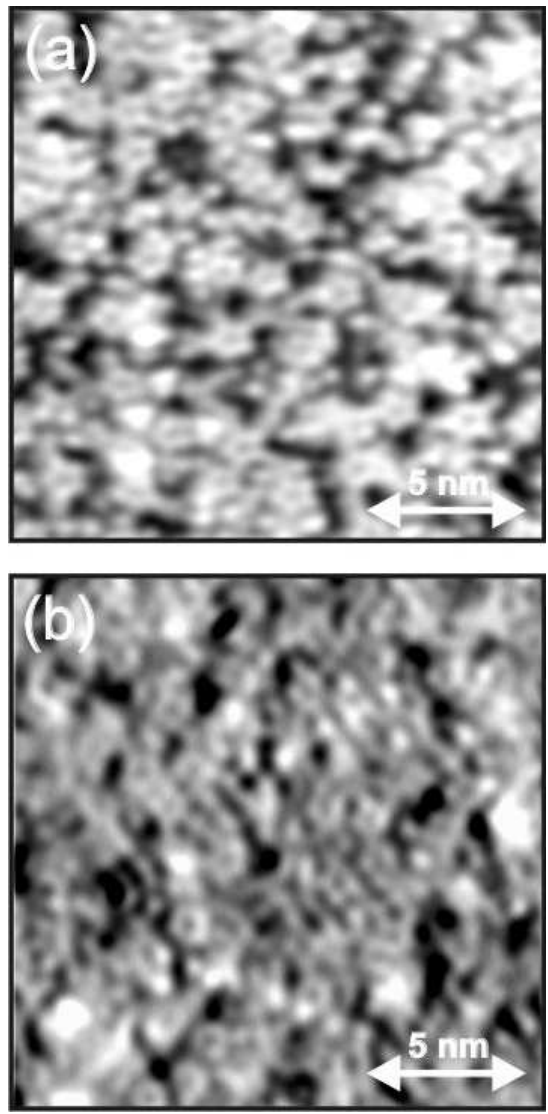}
\caption{Overview of the nucleation and growth of the Bi monolayer by room temperature deposition  on \emph{i}-Al-Pd-Mn. (\emph{a}); 0.54 ML Bi; (\emph{b}); 0.9 ML Bi.}
\label{bigrowth2}
\end{center}
\end{figure}

In the deposition process, the saturation island density corresponds to the point at which the mean free path of the diffusing adatoms is equal to the mean island separation and adatoms will attach themselves with much higher probability to existing islands than to create new ones. The saturation island density is usually reached at coverages of about half a monolayer \cite{Brune98}, and this is the case in the present system: although Bi clusters are still visible,  Fig. \ref{bigrowth2}(a) shows that by 0.54 ML coalescence of islands has begun. This corresponds to the transition from nucleation to growth, and thereafter the system enters the \emph{pure growth regime} \cite{Brune98}. At coverages higher than 0.5 ML, the structural quality of the film is less discernible using STM. Fig. \ref{bigrowth2}({b}) corresponds to a coverage of 0.9 ML. There is still evidence of ring-like structures which we interpret to be Bi clsuters as they have the correct dimensions. However it is no longer possible to confirm their orientation from STM. The FFT from this image still shows quasiperiodic order but the spots are more diffuse than for the lower coverages. At a coverage of 1 ML quasiperiodic order is no longer detectable using STM; we attribute this to the roughness of the film rather than the absence of order. The monolayer film was found to be well-ordered using helium atom scattering (HAS) \cite{Franke02}, although the deposition conditions were slightly different in that study.

Given  the HAS observation of a quasiperiodic monolayer \cite{Franke02}, and the FFT analyses referred to above, the Bi adatoms must nucleate at sites which are quasiperiodically related. There must also be a high density of such sites on the surface - otherwise quasiperiodic ordering at low coverages will not translate to quasiperiodic monolayer growth as was the case with Al-Pd-Mn/Si \cite{Ledieu06} and Al-Pd-Mn/C$_{60}$ \cite{Ledieu01-SS}. When all such sites are occupied, the remaining gaps in the overlayer can be filled with individual Bi atoms, at less energetically favourable sites, leading to the completion of the quasiperiodic monolayer. In the following section we address the identification of the nucleation sites in the pure nucleation regime.

\section{Analysis} \label{analysis}
 The recent application of DFT methodology by Kraj\u{c}\'{\i} and Hafner \cite{Krajci05a} has led to a significantly enhanced understanding of the local structure of the $i$-Al-Pd-Mn surface as imaged using STM. Starting with the Katz-Gratias-Boudard model of the bulk $i$-Al-Pd-Mn quasicrystal structure \cite{Katz93,Boudard92}, they produced bulk approximants of finite unit cell size. The 2/1, 3/2 and 5/3 unit cells of the approximants constructed in this way contain 544, 2292 and 9700 atoms per unit cell respectively. These bulk approximants were then cleaved along certain planes perpendicular to the five-fold axis to produce surface models consisting of several layer slabs of increasing size and complexity. The cleavage planes were chosen to produce planes of high atomic density close to that reported in experiments \cite{Gierer97}.

\begin{figure}
\begin{center}
\includegraphics[width=0.4\textwidth]{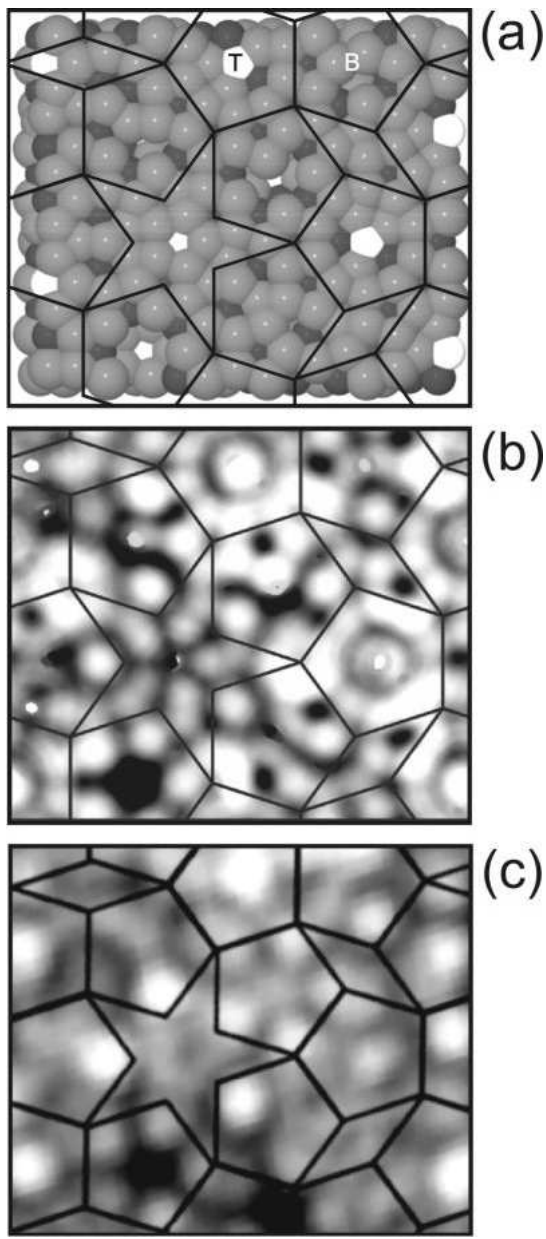}
\caption{(a) The 3/2 approximant surface used by  Kraj\u{c}\'{\i} and Hafner \cite{Krajci05} to model the surface structure of the five-fold surface of $i$-Al-Pd-Mn. Al atoms are shown in light gray; Pd atoms are dark gray and Mn atoms are shown in white. Superimposed is a Penrose P1 tiling of edge length 7.76 \AA{}. $M$ clusters are encompassed by pentagonal tiles while $B$ clusters are centred on the vertices of the tiling. Single `top' and `bottom' tiles have been labelled with `T' and `B' respectively. (b) Simulated STM image of the clean surface with the Penrose P1 tiling superimposed, reproduced from ref. \cite{Krajci06c}. `Top' pentagons with a bright protrusion at the centre correspond to truncations of $M$ clusters with Mn atoms in the surface plane. `Bottom' pentagons sometimes encompass five-fold hollows which Kraj\u{c}\'{\i} and Hafner have shown are produced by the truncation of  $M$ clusters at 2.56 \AA{} above their centre.  (c) STM data from a 40 \AA{} $\times$ 40 \AA{} patch of the clean surface. There is a close match to the calculated patch in (b). `Top' pentagons containing a Mn atom have a bright protrusion at their centre.}
\label{3-2composite}
\end{center}
\end{figure}

Their approach is illustrated in Fig. \ref{3-2composite}. In Fig. \ref{3-2composite}(a), a Penrose P1 tiling  of edge-length 7.8 \AA{} is superimposed on a section of the 3/2 approximant surface \cite{Krajci06c}.  The bulk structure of the $i$-Al-Pd-Mn quasicrystal can be interpreted as being composed of pseudo-Mackay and Bergman clusters \cite{Gratias00}. In the nomenclature used by Gratias et al. \cite{Gratias00} and Kraj\u{c}\'{\i} and Hafner \cite{Krajci05} these are labelled $M$ and B clusters respectively. $B$ clusters are centred on the vertices of the tiling  (though not every vertex corresponds to such a cluster as the minimum distance between them in the bulk is 7.76 \AA{}). The pentagons of the P1 tiling contain the $M$ clusters. There are two different cuts through $M$ clusters. $M$ clusters have a Mn atom at their centre, and when these Mn atoms are in the top-most layer, one orientation of the pentagonal tiles contains these atoms. These tiles are labelled `top' tiles following the notation of Kraj\u{c}\'{\i} and Hafner \cite{Krajci05}. A pentagonal tile of the opposite orientation corresponds to a different cut through an $M$ cluster, where the central Mn atom is 2.56 \AA{} below the surface. These are labelled `bottom' pentagons.  Examples of both `top' and `bottom' pentagons are indicated on Fig. \ref{3-2composite}. Adjacent `top' and `bottom' pentagons always share an edge. The distance between the centres of two adjacent `top' and `bottom' pentagonal tiles is 10.73 \AA{}. On the other hand adjacent pentagonal tiles of the same orientation can only share a vertex; the distance between their centres is 12.62 \AA{}.

Fig. \ref{3-2composite}(b) shows a previously published STM image \cite{Krajci06c} simulated under constant current conditions utilising the surface charge density distribution of the 3/2 approximant model and using the Tersoff-Hamann approximation \cite{Tersoff85,Lin97}. The Mn atoms image brightly in STM, and these features have previously been labeled as `white flowers' in other studies \cite{Papadopolos02}. The `white flower' $M$ clusters in STM are contained in `top' pentagons only. There is some variation in the brightness of the Mn atoms because of a variation in their magnetic state \cite{Krajci06c}. The `dark stars' are always contained in `bottom' pentagons. Due to the irregular filling of the first dodecahedral shell around the Mn atoms, not all of the `bottom' pentagons are equivalent at the surface. Some give rise to well-defined `dark stars' on the surface (see Fig. \ref{3-2composite}(b)). In many cases these `bottom' tiles contain extra atoms from the first irregular shell surrounding the centre of the cluster and the `dark star' is obscured or not observed at all. Fig. \ref{3-2composite}(c) shows a 40 \AA{} $\times$ 40 \AA{} STM image of the clean surface, chosen to be a close match structurally to the section of the model shown in Fig. \ref{3-2composite}(a) and the simulated STM images shown in Fig. \ref{3-2composite}((b). The identification of specific features is clearly facilitated by this approach.

\begin{figure}
\begin{center}
\includegraphics[width=0.4\textwidth]{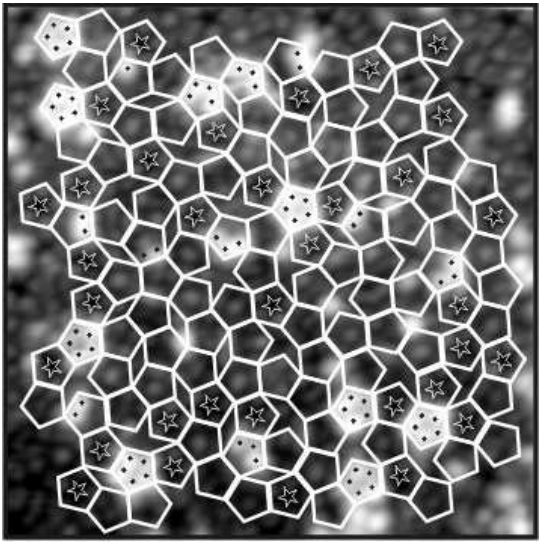}
\caption{The five-fold surface of $i$-Al-Pd-Mn with a Bi coverage of 0.13 ML as in Fig. \ref{bigrowth}(a). A Penrose P1 tiling edge length 7.8 \AA{} has been superimposed.  Bismuth atoms only decorate `top' pentagonal tiles and many of the `bottom' pentagonal tiles enclose a five-fold depression (`dark star'). These are indicated with a star motif.}
\label{tiling}
\end{center}
\end{figure}

The nucleation site for the pentagonal Bi clusters can be understood in terms of this structural framework. The clear visibility of the `dark stars' in the image shown in Fig. \ref{bigrowth} suggests a pathway for nucleation site identification. In Fig. \ref{tiling}, a Penrose P1 tiling has been constructed on an STM image of the 0.13 ML coverage of Bi on $i$-Al-Pd-Mn using the `dark stars' as a guide. In this tiling, every `dark star' falls in a pentagon of a single orientation. According to the description of the clean surface described above, these are the `bottom' tiles. Apart from one or two single Bi atoms, every cluster or partial cluster falls in a pentagon of the opposite orientation. These must therefore occupy the `top' tiles, i.e. the tiles with a Mn atom in the surface plane. We note also that the distance measured between the centres of adjacent Bi pentagons is 12.0 $\pm$ 1.0 \AA{}; this is consistent with the expected distance from the 3/2 model described above.

\begin{figure}
\begin{center}
\includegraphics[width=0.4\textwidth]{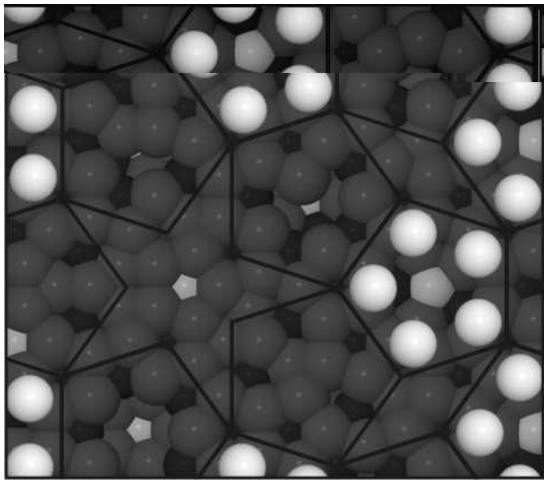}
\caption{Initial nucleation sites of the Bi clusters. As for Fig. \ref{3-2composite}(a), Al atoms are shown in light gray; Pd atoms are dark gray and Mn atoms are shown in white.The substrate atoms are shaded as indicated in the caption for Fig. \ref{3-2composite}. Bi atoms are shown in off-white.}
\label{bipentagons}
\end{center}
\end{figure}

 The number density of `top' pentagons in an infinite Penrose tiling is $\sim$36\% \cite{Baake90}. The suggested initial adsorption sites are shown in Fig. \ref{bipentagons}. The Bi-Bi distance is 4.9 \AA{}. This is a factor $\tau^{-1}$ smaller than the edge-length of the tiling which is 7.76 \AA{}. This geometry suggests that each Bi atom sits in a quasi-three-fold adsorption site. The majority of these sites contain only Al atoms, although a significant number contain two Al atoms and one Pd atom. The height of the Bi atoms above the surface can be roughly estimated from the line profile analysis as 1.2 $\pm$ 0.2 \AA{}. From this estimate, the distance of each Bi atom to the central Mn atom is $\sim$ 3.9 \AA{}. Therefore it is likely therefore that the influence of the Mn atom is felt indirectly through a substrate-mediated interaction.

The relative frequencies of the pentagon, rhomb, star and boat tiles in the infinite Penrose P1 tiling are respectively \cite{Baake90}:
\begin{equation}
1:\frac{1}{\tau^4}:\frac{2\tau-1}{5\tau^5}:\frac{1}{\tau^5}
\end{equation}
which corresponds to percentage occurrences of 72.4:14.6:4.0:9.0 respectively. In the tiling patch in Fig. \ref{tiling}, the corresponding percentage occurrences are 72.6:14.5:0.9:12.0. The close match in these relative frequencies indicates that this tiling patch  is large enough to be representative of the surface structure.

\section{Discussion} \label{discussion}

The Bi atoms are found to adsorb initially in pentagonal clusters with an edge-length of 4.9 \AA{}, in the interior of the `top' pentagons which enclose a truncated pseudo-Mackay cluster with a Mn atom in the surface plane. This finding does not agree with the predictions of the DFT calculations. As described in Section \ref{intro}, the most favourable adsorption sites found by the DFT study, in order of decreasing binding energy, were the surface vacancies or five-fold hollows (-4.9 eV), atop Mn atoms (-4.58 eV), at the vertices of the tiling (-4.37-- -4.32 eV), at the mid-edge\"{y} positions of the tiling (-3.92 eV) and only then the formation of pentagonal Bi clusters in the interior of the pentagons of the tiling. There are several possible reasons for this discrepancy. Firstly, the DFT calculations assumed single Bi atoms at the above adsorption sites. This neglects any interatomic interactions between Bi atoms which might stabilise them in pentagonal clusters. Secondly the calculations were performed at zero temperature. Thus the atoms are static and possible small reconstruction effects within the P1 tiles which might optimize the adsorption site symmetry are not considered. Finally there could be effects due to the small size of the 2/1 model slab used in the calculations. However such calculations on the 3/2 or 5/3 approximant models may not be feasible due to the large number of atoms involved.

The observation of  Bi pentagons with a single orientation, even at higher coverages, indicates that the film should have predominantly five-fold symmetry. This is in agreement with the LEED observations of Franke et al. for the monolayer coverage \cite{Franke02}, and with scans of the He diffraction intensity along high symmetry directions presented in the same paper. We note also that again this is in contrast to the predictions of the DFT results, where the proposed quasiperiodic overlayer  was found to have `pseudodecagonal' symmetry \cite{Krajci05,Krajci06a}.

The analysis rests on the interpretation of the STM data in comparison to the model calculations and simulations carried out by Kraj\u{c}\'{\i} and Hafner \cite{Krajci06c}. Although they used a surface termination which provides an atomic density and composition consistent with other experimental work \cite{Gierer97}, they point out that other possible terminations might produce features similar to those simulated in their study. In particular they note that certain truncations of $B$ clusters which have been previously proposed as possible origins of the `dark star' features \cite{Papadopolos02,Ledieu03} are not included in the chosen termination. Notwithstanding this caveat, the close match of the actual and simulated STM images in ref. \cite{Krajci06c} and again in Fig. \ref{3-2composite} must be considered as an experimental validation of this approach.

These results throw some new light on the conditions under which a quasiperiodic monolayer can be produced. For  Al-Cu-Fe/Al \cite{Cai03} and Al-Pd-Mn/C$_{60}$ \cite{Ledieu01-SS} adsorption was found to take place in the five-fold hollow positions. As noted above, there are not enough of these sites on the surface to present a critical mass and allow the construction of a quasiperiodic framework. Although Si atoms do adsorb in the centres of $M$ clusters in `top' tiles \cite{Ledieu06}, they adsorb not in pentagonal clusters but as individual atoms. The distance between the centres of neighbouring $M$ `top' clusters is 12.62 \AA{}, so that a dispersed network of Si results. When all of the dispersed sites are occupied, there is not a sufficiently dense framework in place to force quasiperiodic order on any further adsorbates.

\section{Conclusions} \label{conclusions}

Bismuth has been deposited onto icosahedral Al-Pd-Mn and its growth studied in the sub-monolayer regime. Preferential adsorption is observed for pentagonal Bi clusters of edge length 4.9 $\pm$ 0.2 \AA~ centred on `top' tiles in the P1 tiling.  Although previous \emph{ab initio} calculations are correct in predicting the adoption of a pseudomorphic growth mode for this system, the initial adsorption site is not that predicted using DFT.  The current observations support the initial He-diffraction study in identifying the symmetry present as five-fold.

\section{Acknowledgments}

The EPSRC is acknowledged for funding this project under Grant number EP/D05253X/1.  Julian Ledieu and Vincent Fourn\'{e}e are thanked for valuable discussions regarding the Pb film. Ed Boughton is acknowledged for assistance with data acquisition, and Kirsty Young is thanked for calculations of tile frequencies. Marian Kraj\u{c}\'{\i} is thanked for providing the Al-Pd-Mn  approximant surface structure models and for useful discussions.


\end{document}